\documentclass[11pt]{article}

\newcommand{\mref}[1]{(\ref{#1})}
\newcommand{\Hami}{{\mathcal H}_J\{\sigma\}}
\newcommand{\ami}[1]{A_J\{\sigma^{#1}\}}
\newcommand{\defeq}{\stackrel{{\mathrm d \mathrm e \mathrm f}}{=}}
\newcommand{\me}[1]{\langle{#1}\rangle}
\newcommand{\Me}[1]{\left\langle{#1}\right\rangle}
\newcommand{\unme}{\frac{1}{2}}
\newcommand{\unte}{\frac{1}{3}}
\newcommand{\unsu}[1]{\frac{1}{#1}}
\newcommand{\EE}{{\mathrm I} \hspace{-0.21 em} {\mathrm E}}   
\newcommand{\cond}[1]{\EE\left(#1|{\mathcal A}_{s}\right)}

\begin{document}

\title{General properties of overlap probability distributions in
disordered spin systems. \\Toward Parisi ultrametricity.}
\date{}
\author{Stefano Ghirlanda\small{*} \and Francesco Guerra\small{**}} 
\maketitle
\centerline{\small{* Zoologiska Institutionen, Stockholms Universtitet, S-106 91 Stockholm, Sweden}}
\centerline{\small{** Dipartimento di Fisica, Universit\`a di Roma ``La Sapienza'', I-00185 Roma, Italy}}
\centerline{\small{\& INFN, Sezione di Roma 1.}}

\begin{abstract}
For a very general class of probability distributions in disordered 
Ising spin systems, in the thermodynamical limit, we prove the following
property for overlaps among real replicas. Consider the overlaps among $s$ replicas. Add one replica $s+1$. Then, the overlap $q_{a,s+1}$ between one of the first $s$ replicas, let us say $a$, and the added $s+1$ is either independent of the former ones, or it is identical to one of the overlaps $q_{a b}$, with $b$ running among the first $s$ replicas, excluding $a$. Each of these cases has equal probability $1/s$.    
\end{abstract}

\section{Introduction}
In this paper we focus on general properties of overlap distributions
in statistical-mechanical models made up of Ising spins (see below for
definitions). Historically,
these properties have been considered for the first time in spin-glass
models \cite{spi87}, so that for convenience we take them as a starting point in
our discussion, and generalise our results later on.

The problem of finding the phase
structure of short-ranged models for spin glasses has proved extremely
difficult, and yet remains unsolved. An important result, though, has
been achieved with Parisi's solution of the Sherrring\-ton-Kirkpatrick
(SK) model (a mean-field 
approximation to more realistic ones), whose
hamiltonian we recall:
\begin{equation}\label{HSK}
{\mathcal H}_J\{\sigma\} = -\unsu{\sqrt{N}}\sum_{(ik)} J_{ik}
\sigma_i \sigma_k\,.
\end{equation}
The $\sigma_i$'s ($i=1,\ldots,N$) are Ising spins and the $J_{ik}$'s
(collectively noted $J$) are  random variables drawn from
independent unit normal distributions, with the constraints $J_{ik}=J_{ki}$ and
$J_{ii}=0$. 
The sum runs over all couples $(ik)$, with $1\le i<k\le N$.
Parisi solution for this model implies a countable infinity of pure
states (below the critical temperature), that turn out to be organised
in a very remarkable geometric structure, of the type called
ultrametric \cite{spi87}.

This structure is clearly seen by introducing a replicated system,
made up of non-interacting, identical copies (replicas) of an SK
system. These are usually called ``real'' replicas to distinguish them
from the replicas used in the replica method,
 which requires a limit to zero replicas. In this way the Boltzmann state of the replicated system is
simply the following product state:
\begin{equation}
\Omega_J(\cdot)=(\omega_J^{(1)}\otimes\cdots\otimes\omega_J^{(s)})(\cdot)\,,
\end{equation}
where $\omega_J^{(i)}$ is the state of replica $i$ for a given
realisation of the $J$'s. We call $\sigma^a_i$ the variables associated to replica $a$, on which $\omega_J^{(a)}$ acts. Notice that all replicas have the \emph{same} noise $J$. We will write
$\EE(\cdot)$ the average over the coupling distribution.

The overlap between the spin configurations of different replicas $a$ and $b$ is
defined as:
\begin{equation}
Q_{ab}=\unsu{N}\sum_{i=1}^N\sigma_i^a\sigma_i^b\,.
\end{equation}
Notice that $Q_{aa}$ would be trivially equal to one.
We introduce now the random variables $q_{ab}$ 
(referred to also as
overlaps) by requiring that for any smooth
function $F_s(Q)$ of the configuration overlaps among $s$ replicas
(summarised by $Q$ in the notation) the following equality holds:
\begin{equation}\label{distrq}
\EE(\Omega_J(F_s(Q)))=\me{F_s(q)}\,,
\end{equation}
where $\me{\cdot}$ is by definition the average with respect to the distribution of
the $q$'s. According to \mref{distrq}, this implies both Boltzmann and
$J$ averages.  Although the replicas are independent so long as the
thermal average is considered, they are coupled by the $J$ average, for
they all share the same realisation of the couplings.  In
fact the Parisi solution gives the following joint distribution for
the  overlaps $q_{12}$ and $q_{13}$ among three replicas, labelled $1$,
$2$, and $3$:
\begin{equation}\label{ro1213}
\rho_{12,13} (q_{12},q_{13})=\unme\rho(q_{12})\rho(q_{13})+\unme\rho(q_{12})\delta
(q_{12}-q_{13})\,,
\end{equation}
where $\rho(\cdot)$ is the probability distribution of the overlap between
any two replicas. Formula \mref{ro1213} says that the two overlaps are
independent with probability one half, and identical with the same
probability.  Even when we consider two overlaps between two distinct
couples of replicas, the correlation remains strong:
\begin{equation}\label{ro1234}
\rho_{12,34}(q_{12},q_{34})=\frac{2}{3}\rho(q_{12})\rho(q_{34})+\unte\rho(q_{12})\delta
(q_{12}-q_{34})\,.
\end{equation}
The difference with formula \mref{ro1213} is just that the probability
of the two overlaps being identical has reduced to one third, and
accordingly the probability of their being independent is two thirdies.

In general, the underling ultrametric structure allows all joint
probability distributions of the overlaps among replicas to be
expressed as functions of the distribution $\rho(\cdot)$ of a single
overlap. We refer to the literature \cite{spi87} for a thorough exposition of
this geometrical structure, but here we want to stress the well known
fact that this situation requires $\rho_J(\cdot)$, i.e. the overlap
distribution at fixed $J$, to be non-self-averaging with respect to
$J$, that is to depend on the realisation of the couplings even after
the thermodynamical  limit has been taken.  In fact, since
$\rho(q)=\EE(\rho_J(Q))$:
\begin{eqnarray}\nonumber
&&\rho_{12,34}(q_{12},q_{34})=\EE(\rho{_J}_{12,34}(Q_{12},Q_{34}))\\&&=\EE(\rho_J(Q_{12})\rho_J(Q_{34}))
\neq\EE(\rho_J(Q_{12}))
\EE(\rho_J(Q_{34}))\,,
\end{eqnarray}
i.e. $\rho_J(\cdot)$ fluctuates with $J$.

While a complete mathematical proof of the Parisi solution is still
lacking, the latter is widely believed to be correct due to extensive
computer simulations \cite{mar96,ini97} and mathematical arguments, 
see \cite{gue96,pas92,tal98,bov97} and references therein.  On the other
hand, doubts about the relevance of the ultrametric structure to
realistic models have been raised in the last years, and in particular
the existence of non-self-averaging quantities in the thermodynamical
limit has been questioned \cite{new96}, see however \cite{
par96}.  In our opinion, since Parisi's method seems very difficult to generalise beyond the mean-field
approximation, a deeper understanding of the ultrametric geometry in
the simpler SK model is a necessary first step in order to consider
its relevance to short-ranged models.  In this paper, following the
ideas outlined in \cite{gue96,ghi96}, we are able to prove that some features
of the ultrametric geometry arise naturally in the SK
model as a consequence of the self-averaging
of thermodynamical quantities, e.g.~the internal energy. 

Moreover, by a careful limiting procedure we establish the same
results for short-ranged models. We do not
address the very difficult question of the number of phases in the
glassy regime of spin-glass models, but show that, whichever this number
be, the overlap distributions (as previously defined) have some ultrametric 
features. 

In section two we recall some known results, and find their
consequences for the overlap distributions in the SK model. In section
three we generalise these arguments in order to construct a
well-defined set of such distributions, independently from Parisi's
work and in agreement with it. Section four is devoted to the
extension of these results to short-ranged models. A short concluding
section summarises our arguments. 

\section{Consequences of self-averaging}
For the sake of convenience let us define:
\begin{equation}
\ami{}\defeq - \unsu{N}\Hami\,.
\end{equation}
We start our arguments by recalling a known theorem, stating the
self-averaging of $\ami{}$ with respect to the $\me{\cdot}$ average in
the thermodynamical limit \cite{gue96,ghi96}:
\begin{equation}\label{automedia}
\lim_{N\rightarrow\infty}\left(\langle\ami{}^2\rangle-\me{\ami{}}^2\right)=0\,.
\end{equation}

By following \cite{gue96,ghi96}, let us sketch the proof of \mref{automedia}. Let us write the $\me{\cdot}$ mean square deviation in \mref{automedia} as a sum of two terms in the form
\begin{eqnarray}
\nonumber\langle A_J\{\sigma\}^2\rangle - \langle A_J\{\sigma\}\rangle^2 = \EE\omega_J(\ami{}^2) - (\EE\omega_J(\ami{}))^2 = \\
\label{autoproof}
\EE(\omega_J(\ami{}^2)-\omega_J^2(\ami{})) + (\EE\omega_J^2(\ami{}) - (\EE\omega_J(\ami{}))^2).
\end{eqnarray}
The second term is the $\EE$ mean square deviation of the internal energy. Since Pastur and Scherbina \cite{pas92} have proven the self-averaging of the free
 energy,
standard thermodynamical reasoning based on convexity gives also self-averaging for the internal energy \cite{gue96,shc97}. This may fail in principle only for a zero measure set of values for $\beta$. Due to the lack of complete control on the thermodynamical limit, it is also necessary to exploit subsequences, as explained in \cite{gue96}. Therefore, the second term in \mref{autoproof} goes to zero as $N\to\infty$. The first term is equal to $N^{-1}\partial_{\beta}\EE\omega_J(\ami{})$. Since  
$\EE\omega_J(\ami{})$ is finite, the $N^{-1}$ term forces also the first term in \mref{autoproof}
to go to zero as $N\rightarrow\infty$, with the possible exclusion of a set of zero measure of values for $\beta$. Therefore we have established \mref{automedia}.     
This, in turn, implies that:
\begin{equation}\label{fluctAF}
\lim_{N\rightarrow\infty}\left(\me{\ami{a}F_s(q)}-\me{\ami{}}\me{F_s(q)}\right)=0\,,
\end{equation}
where by $\ami{a}$ we intend that $\ami{}$ is calculated on replica
$a$, that we take to be one of the replicas in $F_s(q)$. This result
is achieved from
\mref{automedia} via a simple 
Schwarz inequality:
\begin{eqnarray}\nonumber
\lim_{N\rightarrow\infty}\left(\me{\ami{a}F_s(q)}-\me{\ami{}}
\me{F_s(q)}\right)^2 &=& \\ 
\nonumber\lim_{N\rightarrow\infty}\me{\left(\ami{a}-\me{\ami{}}
\right)F_s(q)}^2 &\leq& \\
\lim_{N\rightarrow\infty}\me{(\ami{a}-\me{\ami{}})}^2\me{F_s(q)}^2 &=&
0\,, 
\end{eqnarray}
by virtue of \mref{automedia}. We express now \mref{fluctAF} in terms
of overlaps, using the fact that integration by parts gives
$\me{J_{ik}f(J)}=\me{\partial_{J_{ik}}f(J)}$, for 
a generic function $f(\cdot)$, since the $J$'s are normally
distributed. From this formula we get:
\begin{equation}
\EE\left(\partial_{J_{ik}}\Omega_J(F_s(Q))\right)=\unsu{\sqrt{N}}\sum_{a=1}^s
\EE\left(\Omega_J(F_s(Q)\sigma_i^a\sigma_k^a)-\Omega_J(F_s(Q))
\Omega_J(\sigma_i^a\sigma_k^a)\right)
\end{equation}
by explicit calculation. Applying these formulae, the first term in
\mref{fluctAF} is written as:
\begin{eqnarray}\nonumber
\me{\ami{a}F_s(Q)}&=&\frac{\beta}{N^2}\sum_{(ik)}
\EE\left(\Omega_J\left( \sigma_i^a \sigma_k^a F_s(Q)\sum_{b=1}^s \sigma_i^b
\sigma_k^b\right) -\right.\\ \nonumber
&&\left.\Omega_J\left( \sigma_i^a \sigma_k^a
F_s(Q)\right)\Omega_J\left(
\sum_{b=1}^s  \sigma_i^b \sigma_k^b\right)\right)\,\\
\label{firsthalf} &=&\frac{\beta}{2}\left\langle\left(
\sum_{b=1}^s q_{ab}^2 -sq_{a,s+1}^2\right)F_s(q)\right\rangle\,,
\end{eqnarray}
while the second one is simply:
\begin{equation}\label{secondhalf}
\me{\ami{}}\me{F_s(q)}=\frac{\beta}{2}\left(1-\me{q^2}\right)\me{F_s(q)}\,.
\end{equation}
Using \mref{firsthalf} and \mref{secondhalf} in \mref{fluctAF} we get:
\begin{equation}\label{together}
\lim_{N\rightarrow\infty}\Me{F_s(q)\left(\sum_{b=1}^sq_{ab}^2-sq_{a,s+1}^2-\left(1-\me{q^2}\right)\right)}=0\,.
\end{equation}
Since $F_s(\cdot)$ is a generic function, we can introduce conditional
expectations $\cond{\cdot}$ with respect to the algebra ${\mathcal A}_s$ generated by
the overlaps among $s$ replicas, and write:
\begin{equation}\label{Eas+1}
\cond{q_{a,s+1}^2}=\unsu{s}\me{q^2}+\unsu{s}\sum_{b\neq a}q_{ab}^2\,,
\end{equation}
where the unity in \mref{together} has canceled with the term $a=b$ in
the sum. We shall leave understood that \mref{Eas+1}, as
others formulae obtained in the following, holds exactly only in the
thermodynamical limit.

Using \mref{Eas+1} we can write equalities relating
averages of squared overlaps, the simplest of which is:
\begin{equation}
\me{q_{12}^2q_{13}^2}=\unme\me{q_{}^4}+\unme\me{q_{}^2}^2\,,
\end{equation}
in full agreement with the Parisi probability distribution
\mref{ro1213}.

Moreover, from \mref{Eas+1}
we can easily derive an expression for the conditioned expectation
$\cond{q_{s+1,s+2}^2}$ as well, starting by writing \mref{Eas+1} in
the case of $s+2$ replicas and $a=s+1$: 
\begin{equation}
\EE\left(q_{s+1,s+2}^2|{\mathcal A}_{s+1}\right)=\unsu{s+1}\me{q^2}+\unsu{s+1}\sum_{b=1}^sq_{b,s+1}^2\,.
\end{equation}
Keeping in mind that $\cond{\EE\left(\cdot|{\mathcal
A}_{s+1}\right)}=\cond{\cdot}$, we have:
\begin{eqnarray}\nonumber
\cond{q_{s+1,s+2}^2} &=& \unsu{s+1}\me{q^2}+\unsu{s+1}\sum_{b=1}^s
\cond{q_{b,s+1}^2} \\ &=& \unsu{s+1}\me{q^2}+ 
\unsu{s+1}\left(\me{q^2}+\unsu{s}\sum_{b=1}^s\sum_{c\neq b}
^{1,s}q_{bc}^2\right)\,,
\end{eqnarray}
that is:
\begin{equation}\label{Es+1s+2}
\cond{q_{s+1,s+2}^2}=\frac{2}{s+1}\me{q^2}+\frac{2}{s(s+1)}\sum_{a<b\leq s}q_{ab}^2\,.
\end{equation}
From \mref{Es+1s+2} we can derive other known ultrametric equalities,
for example:
\begin{equation}
\me{q_{12}^2q_{34}^2}=\unte\me{q_{12}^4}+\frac{2}{3}\me{q_{12}^2}^2\,,
\end{equation}
again in agreement with \mref{ro1234} obtained from Parisi's solution.
In the following section we generalise formulae \mref{Eas+1} and
\mref{Es+1s+2} to arbitrary (integer) powers of overlaps.

\section{Auxiliary interactions and overlap probability distributions}
Let us consider the SK model in the presence of an external field:
\begin{eqnarray}\nonumber
A_{J,J'}\{\sigma\}&\defeq&
A_J\{\sigma\}+\frac{\lambda}{N}\sum_{i=1}^{N}J_i'
\sigma_i \\ &\defeq& A_J\{\sigma\} +\lambda I_{J'}\{\sigma\}\,,
\end{eqnarray}
where the random variables $J_i'$ are independt from the $J_{ik}$'s
and with the same distribution. We assume that $\lambda$ is ``small'',
since in the end we will take it to zero to recover the free SK model.

Theorem \mref{fluctAF} can be generalised to the present case, since
it only relies on self-averaging of the internal energy:
\begin{equation}
\lim_{N\rightarrow\infty}\left(\me{A_{J,J'}\{\sigma^a\}F_s(q)}-\me{A_{J,J'}\{\sigma\}}\me{F_s(q)}\right)=0\,, 
\end{equation}
where now $\me{\cdot}$ implies averaging over the $J'$ variables as
well. Using the same procedure as in the preceding section, but now
integrating and deriving with respect to the $J'$ variables, we get
the completely analogous formula:
\begin{equation}
\cond{q_{a,s+1}}=\unsu{s}\me{q}+\unsu{s}\sum_{b\neq a}q_{ab}\,,\\
\end{equation}
which continues to hold when $\lambda$ is taken to zero (after having
taken the thermodynamical limit). The only difference between this
formula and \mref{Eas+1} is that here overlaps appear at the first
power.

It is now clear that we can consider auxiliary interactions of the
general form:
\begin{equation}\label{moltispin}
\lambda_rI_r\{\sigma\}\defeq \frac{\lambda_r}{N^{(r+1)/2}}\sum_{(i_1\ldots i_r)}
J_{i_1,\ldots,i_r}'\sigma_{i_1}\cdots\sigma_{i_r}\,,
\end{equation}
the former case being $r=1$. So we end up with the formula:
\begin{equation}\label{Er1}
\cond{q_{a,s+1}^r}=\unsu{s}\me{q^r}+\unsu{s}\sum_{b\neq a}q_{ab}^r\,,\\
\end{equation}
valid for the free SK model when $\lambda_r\rightarrow 0$. A similar
formula is valid for $\cond{q_{s+1,s+2}^r}$ as a generalisation of \mref{Es+1s+2}. 
We have thus obtained the main result of this paper:
\newline

{\Large\sc theorem.} \emph{Given the overlaps among $s$ real
replicas, the overlap between one of these and an additional replica
is either independent of the former overlaps or it is identical to one of
them, each of these cases having probability $1/s$:}
\begin{equation}\label{result}
\rho_{a,s+1}(q_{a,s+1}|{\mathcal
A}_s)=\unsu{s}\rho(q_{a,s+1})+\unsu{s}\sum_{b\neq
a}\delta(q_{a,s+1}-q_{ab})\,,
\end{equation}
\emph{where $\rho_{a,s+1}(\cdot|{\mathcal A}_s)$ is the conditioned
distribution of $q_{a,s+1}$ given the overlaps in ${\mathcal
A}_s$.}
 
{\Large\sc proof.} The theorem is proved by \mref{Er1} and the fact that the overlaps are
bounded.
\newline

{\Large\sc corollary:} \emph{The distribution
of $q_{s+1,s+2}$ conditioned to the overlaps in ${\mathcal A}_s$ is
given by:}
\begin{equation}\label{corollary}
\rho_{s+1,s+2}(q_{s+1,s+2}|{\mathcal
A}_s)=\frac{2}{s+1}\rho(q_{s+1,s+2})+\frac{2}{s(s+1)}\sum_{a<b\leq
s}\delta(q_{s+1,s+2}-q_{ab})\,.
\end{equation}

{\Large\sc proof:} The proof is the same as that leading from
\mref{Eas+1} to \mref{Es+1s+2}.
\newline

We shall comment on theorems \mref{result} and \mref{corollary} in the
concluding section. Notice that the general relations found in the previous theorem imply also the constraints on overlap distributions found in \cite{aiz97}.

\section{Extension to short-ranged models}
While we have been dealing with the SK model so far, it is easy to establish 
the same results for short-ranged models by carefully taking the 
thermodynamical limit. Suppose that a given model has $M$ pure states at a 
given temperature, that we label $\omega_i(\cdot)$,
$i=1,\ldots,M$. The value of $M$ is immaterial for our argument  
(but see the concluding section). Each of these states can
be reached in the thermodynamical limit by means of a suitable field
$h_i$, that for disordered systems such as spin-glasses is presently unknown. 

Let us now note ${\mathcal H}\{\sigma\}$ the hamiltonian of the considered 
short-ranged model, and let us add to it both the field $h_i$ and a 
perturbation of the SK type \mref{moltispin} (now written as ${\mathcal H}_{SK}\{ 
\sigma\}$), in such a way that the hamiltonian becomes:
\begin{equation}
{\mathcal H}\{\sigma\}+h_i+\lambda{\mathcal H}_{SK}\{\sigma\}\,,
\end{equation}
where $\lambda$ is a ``small'' parameter. 

If, after the thermodynamical limit, $\lambda$ is taken to zero, the
system is left in the state $\omega_i(\cdot)$ selected by the field
$h_i$, if the SK interaction has been kept ``small enough'' throughout
the whole process, and 
 provided that the considered model is not unstable with
respect to the perturbation ${\mathcal H}_{SK}\{\sigma\}$.

While $\lambda$ is different from zero, the SK interaction enables us
to perform the same calculations as shown in the preceding sections,
and to establish the same relations \mref{result} and \mref{corollary}
relatively to the considered states of the system.

In general, we can state that the basic property of the overlaps, given by theorem \mref{result}, holds for all states that can be reached by adding a small spin glass interaction of the type \mref{moltispin} to the original interaction, by taking the infinite volume limit, and then by removing the added spin-glass fields.

\section{Concluding remarks and outlook}
In the preceding sections we have seen how an analysis of fluctuations has 
contributed a deeper understanding of the structure of the SK 
model, and how the same arguments can be extended to general models in 
statistical mechanics.
Theorem \mref{result} is a strong constraint that all overlap
probability distributions have to satisfy, but it is not the same as
ultrametricity. It appears instead to be the same as the hypothesis of
``replica equivalence'' in the framework of the replica method. As
Parisi has shown \cite{par98}, this hypothesis enables one to express
all joint overlap distributions in terms of only those that refer to a
\emph{complete set} of overlaps among each given number $s$ of replicas
(the simplest case beyond $\rho(\cdot)$ being
$\rho_{12,13,23}(\cdot,\cdot,\cdot)$ for $s=3$). Moreover, it
can be shown that, under the hypothesis of replica equivalence, the
only ultrametric solution is the usual Parisi solution \cite{par98},
 and the same conclusion remains true
if one starts from \mref{result} instead that from replica
equivalence.

The fact that theorems \mref{result} and \mref{corollary} hold for
basically all statistical mechanical models, compels us to remind
that their consequences can be trivial, when a particular model is
considered. This happens, for example, in the high temperature regime, when all overlaps take a constant value.

As stated in the introduction, the question of what case applies to
short-ranged spin glass models still stays open.

\section*{Acknowledgments}
We acknowledge useful conversations with Pierluigi Contucci, Charles Newman, Giorgio Parisi, Leonid Pastur, Masha Shcherbina, Michel Talagrand.

\end{document}